# Electric-field-induced changes of magnetic moments and magnetocrystalline anisotropy in ultrathin cobalt films

Takeshi Kawabe[1,*], Kohei Yoshikawa[1,*], Masahito Tsujikawa[2,3,‡], Takuya Tsukahara[1], Kohei Nawaoka[1], Yoshinori Kotani[4], Kentaro Toyoki[4], Minori Goto[1,5], Motohiro Suzuki[4], Tetsuya Nakamura[4], Masafumi Shirai[2,3], Yoshishige Suzuki[1,5], and Shinji Miwa[1,5,†]

[1]*Graduate School of Engineering Science, Osaka University, Toyonaka, Osaka 560-8531, Japan*

[2]*Research Institute of Electrical Communication, Tohoku University, Sendai, Miyagi 980-8577, Japan*

[3]*Center for Spintronics Research Network (CSRN), Tohoku University, Sendai, Miyagi 980-8577, Japan*

[4]*Japan Synchrotron Radiation Research Institute (JASRI), Sayo, Hyogo 679-5198, Japan*

[5]*Center for Spintronics Research Network (CSRN), Osaka University, Toyonaka, Osaka 560-8531, Japan*

[*]These authors contributed equally to this work

[†]miwa@mp.es.osaka-u.ac.jp

[‡]t-masa@riec.tohoku.ac.jp

In this study, the microscopic origins of the voltage-controlled magnetic anisotropy (VCMA) in 3$d$-ferromagnetic metals are revealed. Using *in-situ* X-ray fluorescence spectroscopy that provides a high quantum efficiency, electric-field-induced changes in orbital magnetic moment and magnetic dipole $T_z$ terms in ultrathin Co films are demonstrated. An orbital magnetic moment difference of $0.013\mu_B$. was generated in the presence of electric fields of ±0.2 V/nm. The VCMA of Co was properly estimated by the induced change in orbital magnetic moment, according to the perturbation theory model. The induced change in magnetic dipole $T_z$ term only slightly contributed to the VCMA in 3$d$-ferromagnetic metals.

Various magnetoelectric effects in ultrathin ferromagnetic metals have been reported, such as the electric-field-induced modifications of magnetocrystalline anisotropy energy (MAE) [1–3], Curie temperature [4,5], exchange bias [6] and antisymmetric exchange interaction [7]. In particular, the electric-field-induced modification of MAE, which is called voltage-controlled magnetic anisotropy (VCMA), in ferromagnetic 3$d$-metals has been intensively studied because of its great potential for enabling the construction of ultralow-power-consumption electric devices [3]. As has been reported, one mechanism to explain the MAE of low-dimensional ferromagnetic 3$d$-metals is the magnetization direction dependence of the orbital angular momentum of their 3$d$-state. This anisotropy in the orbital magnetic moment influences the MAE through spin-orbit interactions [8]. This mechanism is often regarded as the Bruno mechanism. Hence, the VCMA can be explained by the change of the orbital magnetic moment by selective electron/hole doping into electron orbitals of atoms at the interface. Several theoretical studies have reported that the hybridization and/or modulation of the 3$d$-orbitals is important for the VCMA [9–13]. However, an electric-field-induced change of the orbital magnetic moment has not been experimentally confirmed.

It has been experimentally confirmed that an electrochemical reaction, namely, $O^{2-}$ migration, induces a magnetoelectric effect in FeCoO$_x$/MgO [14], FePt/ion-gel [15], Co/Gd$_2$O$_3$ [16,17], and Fe/BaTiO$_3$ [18] systems. Because such an electrochemical reaction requires a thermal activation process and has a limited operating speed in the sub-millisecond range [17], it is hardly the microscopic origin of the aforementioned VCMA in 3$d$-ferromagnetic metals with high-speed (< 0.5 ns) operation [3,19–21].

In addition to this, recent research has shown the significance of the magnetic dipole $T_z$ term in the 5$d$-state to the VCMA in Pt with proximity-induced spin polarization [22]. In a Fe/Pt/MgO system, the $T_z$ term induction, correlating with electric quadrupole induction, induces a VCMA through the second-order perturbation term including the spin-flip process [23]. From the first-principles study, this quadrupole mechanism has a comparable contribution to the VCMA in the Fe/Pt/MgO system as the Bruno mechanism. To the best of our knowledge, however, there has been no report experimentally demonstrating the relative significance between the Bruno and the quadrupole mechanisms in 3$d$-ferromagnetic metals because of a lack of direct observations of the electric-field-induced change of orbital magnetic moment in ferromagnetic metals.

Here, we report direct evidence of the electric-field-induced change of orbital magnetic moment of ultrathin Co films in terms of *in-situ* X-ray magnetic circular dichroism (XMCD) spectroscopy with ultrahigh efficiency. Utilizing an epitaxial Fe/Co/MgO multilayer system, we obtain an induced orbital magnetic moment anisotropy change of 0.013$\mu_B$ in the presence of electric fields of ±0.2 V/nm. Moreover, the VCMA in the system is estimated well by the induced change of the orbital magnetic moment anisotropy in Co in accordance with the Bruno's perturbation theory model [8].

The following three sets of samples were prepared for the studies. The first set of samples were tunnel junctions to characterize the electric-field-induced change of XMCD using the partial fluorescence yield (PFY) method and a silicon drift detector (SDD) with a large solid angle. The device structure is illustrated in Fig. 1. A multilayer consisting of fcc-MgO(001) substrate/fcc-MgO(001) buffer (5 nm)/bcc-V(001) (30 nm)/bcc-Fe(001) (0.4 nm)/Co (0.14 nm)/fcc-MgO(001) barrier (2 nm) was prepared using electron beam deposition in an ultrahigh vacuum. The MgO(001) buffer was employed to block the diffusion of carbon from the surface of the MgO substrate into the V/Fe/Co multilayer [24]. The MgO substrate was annealed at 800 °C for 10 min, and the V layer was post-annealed at 500 °C for 30 min. We confirmed formations of epitaxial and flat interfaces of each layer using the results of reflection high-energy electron diffraction [25]. The 0.14-nm Co, corresponding to a monatomic Co layer, was grown coherently onto the bcc-Fe(001) surface while maintaining its two-dimensional square lattice structure. Then, the Co layer was covered by the fcc-MgO(001) epitaxial layer. Following the removal of the sample from the ultrahigh vacuum, a $SiO_2$ (5 nm)/Cr (2 nm)/Au (5 nm) layer was deposited. Then, the multilayer was patterned into a 160-μm diameter tunnel junction [26]. The second set of samples were continuous films used to characterize the X-ray absorption spectra (XAS) and its XMCD; the total electron yield (TEY) method was employed to calibrate the XMCD obtained by the PFY method. A similar (001)-oriented epitaxial multilayers that consists of MgO substrate/MgO buffer (5 nm)/V (30 nm)/Fe (0.4 nm)/Co (0 nm, 0.14 nm)/MgO (2 nm), without a top $SiO_2$/Cr/Au layer, was prepared. The third set of samples were magnetic tunnel junctions used to characterize the MAE and VCMA. A similar (001)-oriented epitaxial multilayer that consists of MgO substrate/MgO buffer (5 nm)/V (30 nm)/Fe (0.3 nm, 0.4 nm, 0.5 nm)/Co (0 nm, 0.14 nm)/MgO barrier (1.4 nm)/Fe (10 nm)/Au (5 nm) was prepared and patterned into tunnel junctions that have a 2×5 μm$^2$ junction.

The XAS/XMCD measurements were conducted at the soft X-ray beamline, BL25SU of SPring-8. Details of the measurement setup and conditions are reported elsewhere [26–28]. The degree of circular polarization was previously estimated to be 96% [28] and was used as the correcting factor in the sum rule analysis. The XAS with right and left helicities, $\mu_+$ and $\mu_-$, respectively, were recorded. The self-absorption effects in the fluorescent XAS/XMCD were corrected by referring to those recorded by the TEY method, which are less affected by the self-absorption effect [29]. In the estimations of the magnetic moments using the sum rule, the accuracy errors of the physical parameters are approximately 10%. However, the errors that emerge due to the relative changes, depending on the electric field and/or magnetization direction, i.e., precision errors, are much smaller than the accuracy errors and were used to determine the error bars for this study. All the XAS/XMCD measurements were conducted at room temperature.

The inset of Fig. 2(a) depicts the magnetization as a function of the external magnetic field $H$, which was obtained using the XMCD defined as ($\mu_+ - \mu_-$) at the Co $L_3$ edge (778.4 eV) with the PFY method. The magnetic field direction was $\theta = 20°$. From the inset of Fig. 2(a), it can be seen that the magnetization of the Fe/Co layer is easily saturated in the magnetic field direction because the in-plane shape anisotropy energy and perpendicular MAE almost cancel each other out in the Fe/Co layer. Figures 2(a) and 2(b) present the polarization-averaged XAS defined as ($\mu_+ + \mu_-$)/2 and the XMCD spectra around the $L_3$- and $L_2$-edges of the Co using the PFY method, respectively. Magnetic fields of ±1.9 T were applied at $\theta = 20°$ to saturate the magnetization of the Fe/Co layer. The instrumental asymmetries of the nonmagnetic origin were removed by measuring the spectra with an opposite magnetic field direction.

The magnetic properties of the V/Fe (0.4 nm)/Co (0.14 nm)/MgO (2 nm) and V/Fe (0.4 nm)/MgO (2 nm) are summarized in Table 1, where $m_L$, $m_S$, $m_T$, and $\mu_B$ are the orbital magnetic moment, spin magnetic moment, magnetic dipole moment, and Bohr magneton, respectively. Each magnetic moment was characterized by the XAS/XMCD results obtained using the TEY method with sum rule analysis. For the sum rule analysis [30–32], 2.29 and 3.39 were employed as the number of holes in the $3d$-orbitals of Co and Fe, respectively [32]. The magnetic moments of both Co and Fe in the V/Fe/Co/MgO sample were larger than the reported values for pure Co and Fe [32], while those of Fe in the V/Fe/MgO sample were comparable to the values of pure Fe. Such large magnetic moments were reported in a Fe–Co alloy [33] and Co monatomic layer [34–36]. The interfacial MAE is defined as the MAE with no bulk-induced effects [2], and was characterized by the resonant field of the ferromagnetic resonance in the magnetic tunnel junctions [25]. A positive MAE is defined as preferring perpendicular magnetization. The VCMA (fJ/Vm) was characterized by the electric-field-induced shift in the resonant field and is defined as the MAE (mJ/m$^2$) per unit electric field (V/m) in the 2-nm MgO.

It was reported that the (0001)-oriented ultrathin hcp-Co film exhibits orbital magnetic moment anisotropy, which explains the perpendicular MAE [35]. From Table 1, the orbital magnetic moment of the 0.14-nm Co film in our experiment is relatively large, but its anisotropy is negligible. In contrast to the Co, the Fe exhibits an orbital magnetic moment anisotropy of about 20% in both the V/Fe/Co/MgO and V/Fe/MgO samples. Hence, the interfacial MAE of 0.4 mJ/m$^2$ in V/Fe/Co/MgO and 0.5 mJ/m$^2$ in V/Fe/MgO may be attributed to the MAE of the Fe. The VCMA of Fe/Co/MgO (−82 fJ/Vm) is more than twice that of Fe/MgO (−31 fJ/Vm). Because the Co insertion at the Fe–MgO interface significantly increases the VCMA, the Co should be responsible for the VCMA in the system.

The changes in the Co-$L_3$ (778.4 eV) and -$L_2$ (793.8 eV) peak heights of the XMCD signals induced by the electric field were characterized and are displayed in Figs. 3(a) and 3(b), respectively, which were obtained using the PFY method. A magnetic field of ±1.9 T with $\theta$ = 20° was applied during the measurement to saturate the magnetization of Fe/Co layer in the magnetic field direction. In our system, an external voltage of ±3 V corresponds to an electric field of ±0.2 V/nm according to the capacitance model [2]. The positive (negative) voltage induces electrons (holes) at the Co–MgO interface. The solid and dashed lines indicate forward and backward voltage sweeps, respectively. After the measurements, we can confirm that there is no significant change in the XAS/XMCD before and after the measurements, which implies that sample degradation by the electric field application is negligible. The same measurement with a magnetic field of ±1.9 T with $\theta$ = 70° was also conducted. In contrast to the case of Fe/MgO [26,37], where the induced changes of XMCD at the Fe-$L_2$ and -$L_3$ absorption edges were negligible, it should be noted that significant changes were observed at the Co-$L_2$ and -$L_3$ absorption edges in Fe/Co/MgO.

The electric-field-induced changes in the magnetic moments of Co were determined using sum rule analysis [30–32] and are presented in Figs. 3(c) and 3(d). We assumed that the XMCD integrals at the $L_2$- and $L_3$-edges of Co, calculated by the TEY method, were proportional to their peak intensities displayed in Figs. 3(a) and 3(b). In the sum-rule analysis equations (Eqs. (1)−(3) in Ref. 22), hole number of $d$-orbitals is proportional to the XAS integrals at the $L_3$- and $L_2$- edges. However, the magnetic moments are not dependent of the changes of the XAS integrals. Therefore, although XAS and hole number can be modulated by the electric-field, the influence of the induced changes in XAS on the changes in magnetic moments could be neglected. As shown in Fig. 3(c), the $m_L$ of Co with a voltage of −3 V is larger than that corresponding to 3 V. Moreover, the induced change of $m_L$ with $\theta = 20°$ is larger than that with $\theta = 70°$. Our experiment demonstrates that an orbital magnetic moment anisotropy change of $(0.013\pm0.008)\mu_B$ between magnetization angles of $\theta = 20°$ and $70°$ was generated in the presence of electric fields of ±0.2 V/nm. The electric-field-induced change in the effective spin magnetic moment $m_S-7m_T$ is shown in Fig. 3(d). Similar to $m_L$, $m_S-7m_T$ is enhanced at negative voltage application. Moreover, the electric-field-induced change of the magnetic moment is anisotropic. In contrast to $m_T$, it is known that $m_S$ is insensitive to the magnetization direction. Hence, the anisotropic part of the induced change of the magnetic moment is attributed to $m_T$, that is, the magnetic dipole $T_z$ term [23,31,36]. A similar trend in the induced change of $m_S-7m_T$ was reported in Pt with proximity-induced spin polarization [22]. Note that the enhanced $m_S-7m_T$ at a negative bias voltage, which induces holes at the Co–MgO interface, cannot be explained by the electric-field-induced change in $m_S$ because of an electrochemical reaction following oxygen migration. In the case of oxygen migration, hole accumulation decreases the magnetic moments as reported in Co/GdO$_x$ [16].

In Figs. 3(c) and (d), the electric-field-induced changes of the magnetic moments may be nonlinear; however, such nonlinear components are within the error bars. During the XMCD measurement, the sample current (power) density in the tunnel junction was approximately 0.5 mA/cm$^2$ (1.5 mW/cm$^2$). Therefore, the change of the sample temperature would be negligible, i.e., much lower than 1K [38]. Therefore, a sample temperature change is not likely to cause such nonlinear behavior. Moreover, we have confirmed that VCMA in Fe/Co/MgO is linear as reported in our previous studies [39], where a linear electric-field-induced change in orbital magnetic moment is expected. Therefore, in the following discussion, we employed the values at ±3 V, in order to minimize errors in the analysis.

To analyze the MAE, the following equation is employed to address the second-order perturbation of the spin-orbit interaction [23]:

$$\Delta E \cong \frac{\lambda'}{4\mu_B}\left(\Delta m_{L,\downarrow} - \Delta m_{L,\uparrow}\right) - \frac{21}{2\mu_B}\frac{\lambda'^2}{E_{ex}}\Delta m_T. \tag{1}$$

The perpendicular MAE ($\Delta E > 0$ for perpendicular easy axis) is defined as the MAE of the in-plane magnetized film subtracted from that of the perpendicularly magnetized film. Here, $\Delta m_{L,s}(=m_{L,s}^{0°} - m_{L,s}^{90°})$ and $\Delta m_T(=m_T^{0°} - m_T^{90°})$ express the changes in the orbital magnetic moment and magnetic dipole moment between the perpendicularly ($\theta = 0°$) and in-plane ($\theta = 90°$) magnetized films, respectively. Moreover, $\Delta m_{L,\downarrow(\uparrow)}$ represents the contribution from the minority (majority) spin-band. The measured orbital magnetic moment $m_L^\theta$ is equal to $m_{L,\downarrow}^\theta + m_{L,\uparrow}^\theta$, and $\lambda'$ is the effective spin-orbit interaction coefficient in the 3$d$-bands. In contrast to the case of Pt with proximity-induced spin polarization [22], we first assume that we can neglect the second term corresponding to the spin-flip perturbation process between the exchange-split $E_{ex}$ majority and minority spin-bands, insofar as $E_{ex}$ is large and $\lambda'$ is small in the Co. In the case of Co, we can also assume that we can neglect the term related to $m_{L\uparrow}$ [36], because the majority spin-band is almost occupied. Thus, the perpendicular MAE is proportional to the measured changes in the orbital magnetic moment; this relation is known as the Bruno model [8], $\Delta E \approx (\lambda' \Delta m_L)/4\mu_B$.

The $\lambda'$ for the Bruno model is not identical to the spin-orbit interaction coefficient of an atom, because $\lambda'$ depends on the band structure. The $\lambda'$ of the Co for the Bruno model was reported to be 3.3 meV in the Au/Co(3 monolayer, ML)/Au multilayer [36]. If only the interfacial 2ML-Co is responsible for the observed orbital magnetic moment anisotropy, the intrinsic $\lambda'$ for the Co would be 5.0 meV. In our study, using the Bruno model, if $\lambda'$ = 5.0 meV, we obtained that the induced change in perpendicular MAE is $(0.039 \pm 0.023)$ mJ/m$^2$ when the electric field is switched from +0.2 V/nm to −0.2 V/nm; the experimentally obtained $\Delta m_L = (0.017 \pm 0.010)\mu_B$ was used. We employed a simple assumption, $m_L^\theta = m_L^{0°} \cos^2\theta + m_L^{90°} \sin^2\theta$. Table 1 (−82 fJ/Vm) shows that the experimentally obtained induced change in perpendicular MAE of the Fe/Co/MgO structure at ±0.2 V/nm is 0.03 mJ/m$^2$, which is in good agreement with the induced change in perpendicular MAE obtained using the Bruno model. Hence, our experiment provides direct evidence for the application of the Bruno model to the VCMA in the 3$d$-ferromagnetic metals.

From the discussion above, the change of the orbital magnetic moment anisotropy in the Co seems to explain the VCMA. However, the impact of the change of the magnetic dipole $T_z$ term $m_T$ shown in Fig. 3(d), on the VCMA remains to be seen. Figure 4(a) presents a cross-sectional view of our model for the first-principles study. The Fe/Co/MgO multilayer was modeled by a periodic slab supercell with 3ML-Cu, 4ML-V, 3ML-Fe, 2ML-CoFe, 5ML-MgO, and a 26-Å-thick vacuum layer. To reproduce the magnetic properties in Table 1, we did not employ the Fe/Co/MgO multilayer with an ideal Co–MgO interface, but instead used that with intermixed CoFe at the MgO interface. To simplify the model, we assume that the Co coverage and Fe coverage in the atoms at the MgO interface are both 0.5. Details of the computation method are reported elsewhere [22]. For the in-plane lattice constant of the atoms, the value of 0.286 nm, which is identical to that of bulk Fe, was employed. Because of the strong screening effect of metals, the MAE change in the system is dominated by atoms at the MgO interface. Figure 4(a) also depicts the induced change of charge density. The induced change of charge density at an electric field of +0.2 V/nm in the MgO is subtracted from that at −0.2 V/nm. The blue and red regions indicate the hole accumulations and depletions, respectively. From Fig. 4(a), note that the induced change of the charge density in the metals is dominant in the interfacial atoms with the MgO. Table 2 lists the magnetic properties obtained from the first-principles study. The values of the magnetic moments are those of the atoms at the MgO interface. The first-principles study qualitatively reproduces the experimental results listed in Table 1. To discuss the effects of the electric-field-induced $m_L$ and $m_T$ on the VCMA, the induced change of perpendicular MAE from the second-order perturbation to the spin-orbit interaction is calculated directly from the first-principles study [40]. We employed the following equation as the perpendicular MAE [41]:

$$\Delta E_{s's} = \lambda^2 \sum_{o,u} \frac{\left|\langle u,s',\perp|L_z|o,s,\perp\rangle\right|^2 - \left|\langle u,s',//|L_x|o,s,//\rangle\right|^2}{E_{u,s'} - E_{o,s}}. \qquad (2)$$

For Co and Fe, 69.5 and 54.4 meV, which is the spin-orbit coupling of the atoms, are employed for $\lambda$ in Eq. 2, respectively. The values of the electric-field-induced changes (δ) to the perpendicular MAE in the Co and Fe atoms at the MgO interface, arising from the spin-conserved term ($s's$ = ↑↑ or $s's$ = ↓↓) and spin-flip term ($s's$ = ↓↑ or $s's$ = ↑↓), are derived and shown in Fig. 4(b). The MAE at +0.2 V/nm is subtracted from that at −0.2 V/nm. As discussed in previous studies [22,23], the perpendicular MAE from the spin-conserved terms ($\Delta E_{↑↑}$ and $\Delta E_{↓↓}$) in Eq. 2 corresponds to the first terms of Eq. 1: $\Delta m_{L,↓} - \Delta m_{L,↑}$. Further, the perpendicular MAE from the spin-flip terms ($\Delta E_{↓↑}$ and $\Delta E_{↑↓}$) in Eq. 2 corresponds to the second term of Eq. 2: $\Delta m_T$. First, from Fig. 4(b), the electric-field-induced change of the perpendicular MAE in Co is approximately three times larger than that of Fe. Second, the change in the electric-field-induced perpendicular MAE from the spin-flip terms ($\delta E_{↓↑} + \delta E_{↑↓}$) is negligible and that from the spin-conserved terms ($\delta E_{↑↑} + \delta E_{↓↓}$) dominates the MAE change. From these results, note that the induced perpendicular MAE change from the spin-conserved terms of Co dominates the MAE change in the system. In other words, rather than the electric-field-induced change of the magnetic dipole $T_z$ term $\delta m_T$, the induced change of the orbital magnetic moment anisotropy $\delta m_L$ is responsible for the VCMA.

In this study, the magnetic moments of Co were experimentally characterized using *in-situ* XMCD at ±0.2 V/nm external electric fields. With the electric-field-induced changes in magnetic moments in the 3$d$-state of Co, an induced change in orbital magnetic moment of 0.013$\mu_B$ was confirmed. The VCMA in the Co was suitably estimated by the induced change of the orbital magnetic moment anisotropy, according to the Bruno's perturbation theory model. The induced change of the magnetic dipole $T_z$ term of Co was confirmed; however, the first-principles study indicates that the VCMA in the system is determined by the induced change of the orbital magnetic moment anisotropy rather than that of the magnetic dipole $T_z$ term. This study provides new insight into electric-field control of condensed matter.

We thank E. Tamura of Osaka University, T. Nozaki of AIST, and Y. Shiota of Kyoto University for their discussions. This work received support from the ImPACT Program of the Council for Science, Technology and Innovation (Cabinet Office, Government of Japan), and from JSPS KAKENHI (Grant Nos. JP26103002 and JP15H05420). The XAS and XMCD measurements were performed in SPring-8 with the approval of the Japan Synchrotron Radiation Research Institute (Proposal Nos. 2015A0079, 2016B1016, 2017A1012, 2017A1201, and 2017A1869).

1   M. Weisheit, S. Fähler, A. Marty, Y. Souche, C. Poinsignon, and D. Givord, Science **315**, 349 (2007).

2   T. Maruyama, Y. Shiota, T. Nozaki, K. Ohta, N. Toda, M. Mizuguchi, A. Tulapurkar, T. Shinjo, M. Shiraishi, S. Mizukami, Y. Ando, and Y. Suzuki, Nat. Nanotechnol. **4**, 158 (2009).

3   B. Dieny and M. Chshiev, Rev. Mod. Phys. **89**, 025008 (2017).

4   D. Chiba, S. Fukami, K. Shimamura, N. Ishiwata, K. Kobayashi, and T. Ono, Nat. Mater. **10**, 853 (2011).

5   M. Oba, K. Nakamura, T. Akiyama, T. Ito, M. Weinert, and A. J. Freeman, Phys. Rev. Lett. **114**, 107202 (2015).

6   P. Borisov, A. Hoschstrat, X. Chen, W. Kleemann, and C. Binek, Phys. Rev. Lett. **94**, 117203 (2005).

7   K. Nawaoka, S. Miwa, Y. Shiota, N. Mizuochi, and Y. Suzuki, Appl. Phys. Express **8**, 063004 (2015).

8   P. Bruno, Phys. Rev. B **39**, 865(R) (1989).

9   C.-G. Duan, J. P. Velev, R. F. Sabirianov, Z. Zhu, J. Chu, S. S. Jaswal, and E. Y. Tsymbal, Phys. Rev. Lett. **101**, 137201 (2008).

10  M. Tsujikawa and T. Oda, Phys. Rev. Lett. **102**, 247203 (2009).

11  K. Nakamura, R. Shimabukuro, Y. Fujisawa, T. Akiyama, T. Ito, and A. J. Freeman, Phys. Rev. Lett. **102**, 187201 (2009).

12  P. V. Ong, N. Kioussis, D. Odkhuu, P. K. Amiri, K. L. Wang, and G. P. Carman, Phys. Rev. B **92**, 020407(R) (2015).

13  F. Ibrahim, H. X. Yang, A. Hallal, B. Dieny, and M. Chshiev, Phys. Rev. B **93**, 014429 (2016).

14  F. Bonell, Y. T. Takahashi, D. D. Lam, S. Yoshida, Y. Shiota, S. Miwa, T. Nakamura, and Y. Suzuki, Appl. Phys. Lett. **102**, 152401 (2013).

15  K. Leistner, J. Wunderwald, N. Lange, S. Oswald, M. Richter, H. Zhang, L. Schultz, and S. Fähler, Phys. Rev. B **87**, 224411 (2013).

16  C. Bi, Y. Liu, T. Newhouse-Illige, M. Xu, M. Rosales, J. W. Freeland, O. Mryasov, S. Zhang, S. G. E. te Velthuis, and W. G. Wang, Phys. Rev. Lett. **113**, 267202 (2014).

17  U. Bauer, L. Yao, A. J. Tan, P. Agrawal, S. Emori, H. L. Tuller, S. van Dijken, and G. S. D. Beach, Nat. Mater. **14**, 174 (2015).

18  G. Radaelli, D. Petti, E. Plekhanov, I. Fina, P. Torelli, B. R. Salles, M. Cantoni, C. Rinaldi, D. Gutiérrez, G. Panaccione, M. Varela, S. Picozzi, J. Fontcuberta, and R. Bertacco, Nat. Commun. **5**, 3404 (2014).

19  Y. Shiota, T. Nozaki, F. Bonell, S. Murakami, T. Shinjo, and Y. Suzuki, Nat. Mater. **11**, 39 (2012).

20  T. Nozaki, Y. Shiota, S. Miwa, S. Murakami, F. Bonell, S. Ishibashi, H. Kubota, K. Yakushiji, T. Saruya, A. Fukushima, S. Yuasa, T. Shinjo, and Y. Suzuki, Nat. Phys. **8**, 491 (2012).

21  J. Zhu, J. A. Katine, G. E. Rowlands, Y.-J. Chen, Z. Duan, J. G. Alzaate, P. Upadhyaya, J. Langer, P. K. Amiri, K. L. Wang, and I. N. Krivorotov, Phys. Rev. Lett. **108**, 197203 (2012).

22  S. Miwa, M. Suzuki, M. Tsujikawa, K. Matsuda, T. Nozaki, K. Tanaka, T. Tsukahara, K. Nawaoka, M. Goto, Y. Kotani, T. Ohkubo, F. Bonell, E. Tamura, K. Hono, T. Nakamura, M. Shirai, S. Yuasa, and Y. Suzuki, Nat. Commun. **8**, 15848 (2017).

23  G. van der Laan, J. Phys.: Condens. Matter **10**, 3239 (1998).

24  A. Kozioł-Rachwał, T. Nozaki, V. Zayets, H. Kubota, A. Fukushima, S. Yuasa, and Y. Suzuki, J. Appl. Phys. **120**, 085303 (2016).

25  S. Miwa, J. Fujimoto, P. Risius, K. Nawaoka, M. Goto, and Y. Suzuki, Phys. Rev. X **7**, 031018 (2017).

26  T. Tsukahara, T. Kawabe, K. Shimose, T. Furuta, R. Miyakaze, K. Nawaoka, M. Goto, T. Nozaki, S. Yuasa, Y. Kotani, K. Toyoki, M. Suzuki, T. Nakamura, Y. Suzuki, and S. Miwa, Jpn. J. Appl. Phys. **56**, 060304 (2017).

27  T. Nakamura, T. Muro, F. Z. Guo, T. Matsushita, T. Wakita, T. Hirono, Y. Takeuchi, and K. Kobayashi, J. Electron Spectrosc. Relat. Phenom. **144–147**, 1035 (2005).

28  T. Hirono, H. Kimura, T. Muro, Y. Saitoh, and T. Ishikawa, J. Electron Spectrosc. Relat. Phenom. **144–147**, 1097 (2005).

29  L. Tröger, D. Arvanitis, K. Baberschke, H. Michaelis, U. Grimm, and E. Zschech, Phys. Rev. B **46**, 3283 (1992).

30  B. T. Thole, P. Carra, F. Sette, and G. van der Laan, Phys. Rev. Lett. **68**, 1943 (1992).

31  P. Carra, B. T. Thole, M. Altarelli, and X. Wang, Phys. Rev. Lett. **70**, 694 (1993).

32  C. T. Chen, Y. U. Idzerda, H.-J. Lin, N. V. Smith, G. Meigs, E. Chaban, G. H. Ho, E. Pellegrin, and F. Sette, Phys. Rev. Lett. **75**, 152 (1995).

33  K. Schwarz, P. Mohn, P. Blaha, and J. Kübler, J. Phys. F: Met. Phys. **14**, 2659 (1984).

34  P. Gambardella, A. Dallmeyer, K. Maiti, M. C. Malagoli, W. Eberhardt, K. Kern, and C. Carbone, Nature **416**, 301 (2002).

35  D. Weller, J. Stöhr, R. Nakajima, A. Carl, M. G. Samant, C. Chappert, R. Mégy, P. Beauvillain, P. Veillet, and G. A. Held, Phys. Rev. Lett. **75**, 3752 (1995).

36  J. Stöhr and J. Magn. Magn. Mater. **200**, 470 (1999).

37  S. Miwa, K. Matsuda, K. Tanaka, Y. Kotani, M. Goto, T. Nakamura, and Y. Suzuki, Appl. Phys. Lett. **107**, 162402 (2015).

38  A. Sugihara, M. Kodzuka, K. Yakushiji, H. Kubota, S. Yuasa, A. Yamamoto, K Ando, K. Takanashi, T. Ohkubo, K. Hono, and A. Fukushima, Appl. Phys. Express **3**, 065204 (2010).

39  K. Tanaka, S. Miwa, Y. Shiota, N. Mizuochi, T. Shinjo, and Y. Suzuki, Appl. Phys. Express **8**, 073007 (2015).

40  Y. Miura, S. Ozaki, Y. Kuwahara, M. Tsujikawa, K. Abe, and M. Shirai, J. Phys. Condens. Matter **25**, 106005 (2013).

41  D.-S. Wang, R. Wu, and A. J. Freeman, Phys. Rev. B **47**, 14932 (1993).

TABLE I. Experimentally determined magnetic properties of Fe/Co/MgO and Fe/MgO, where $m_L$, $m_S$, $m_T$, $\mu_B$ denote the orbital magnetic moment, spin magnetic moment, magnetic dipole moment, and Bohr magneton, respectively. The magnetic moments are characterized by the XAS/XMCD results obtained using the TEY method with sum rule analysis, and they are the averaged values in the total thickness. In the estimations of the magnetic moments using the sum rule, the accuracy errors of the physical parameters are approximately 10%. However, the errors that emerge due to the relative changes, i.e. precision errors, are much smaller than the accuracy errors and were used to determine the error bars. The interfacial MAE is defined as the MAE with no bulk-induced effects. The VCMA is defined as the change in MAE per unit electric field in the 2-nm MgO.

|  |  | Fe (0.4 nm)/Co (0.14 nm)/MgO | | Fe (0.4 nm)/MgO |
|---|---|---|---|---|
|  |  | Fe | Co | Fe |
| $m_L/\mu_B$ | $\theta = 0°$ | 0.12 ± 0.005 | 0.28 ± 0.005 | 0.10 ± 0.005 |
|  | $\theta = 70°$ | 0.10 ± 0.005 | 0.28 ± 0.005 | 0.08 ± 0.005 |
| $(m_S - 7m_T)/\mu_B$ | $\theta = 0°$ | 2.63 ± 0.01 | 2.28 ± 0.01 | 2.09 ± 0.01 |
|  | $\theta = 70°$ | 2.23 ± 0.01 | 2.19 ± 0.01 | 2.05 ± 0.01 |
| Interfacial MAE | | 0.4 ± 0.05 mJ/m² | | 0.5 ± 0.05 mJ/m² |
| VCMA | | −82 ± 8 fJ/Vm | | −31 ± 3 fJ/Vm |

TABLE II. Calculated magnetic properties of Fe/Co/MgO and Fe/MgO systems. The values of the magnetic moments are those of the atoms at the MgO interface. The perpendicular MAE is defined as the difference in MAEs between the perpendicularly and in-plane magnetized states. The VCMA is defined as the change in MAE per unit electric field in MgO.

|  |  | Fe/Co/MgO | | Fe/MgO |
|---|---|---|---|---|
|  |  | Fe | Co | Fe |
| $m_L/\mu_B$ | $\theta = 0°$ | 0.112 | 0.144 | 0.127 |
|  | $\theta = 90°$ | 0.098 | 0.129 | 0.102 |
| $(m_S - 7m_T)/\mu_B$ | $\theta = 0°$ | 2.814 | 1.869 | 2.976 |
|  | $\theta = 90°$ | 2.637 | 1.670 | 2.782 |
| Perpendicular MAE | | 0.20 mJ/m² | | 0.98 mJ/m² |
| VCMA | | −256 fJ/Vm | | −90 fJ/Vm |

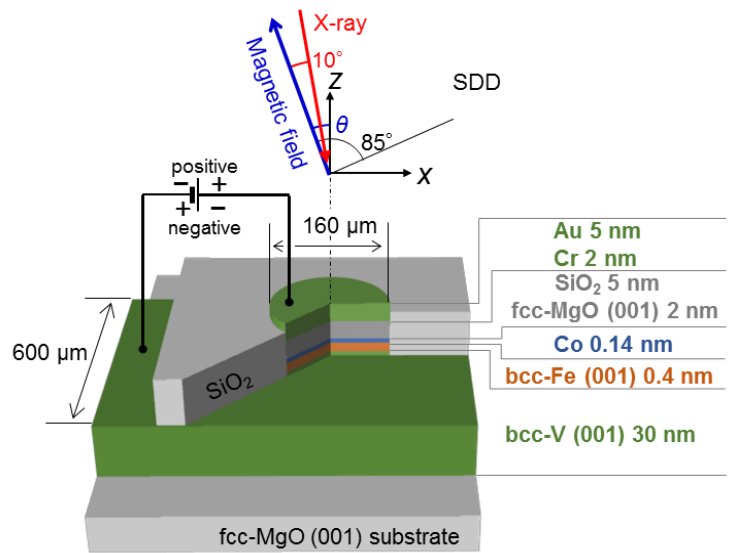

FIG. 1 (Color online) Schematic of the sample structure and measurement configuration. An external electric field was applied to the ferromagnetic ultrathin Co film with a two-dimensional square lattice via a dielectric consisting of MgO and $SiO_2$. XAS/XMCD measurements were performed using the PFY method with an SDD.

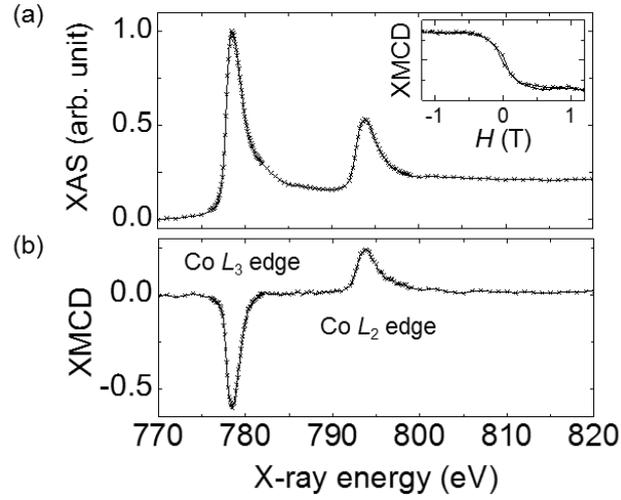

FIG. 2 (a) Polarization-averaged X-ray absorption ($\mu_+ + \mu_-$)/2 and (b) its XMCD ($\mu_+ - \mu_-$) spectra obtained using the PFY method around the Co-absorption edge. An external magnetic field of ±1.9 T was applied to saturate the magnetization of the Fe/Co layer in the magnetic field direction. The inset shows the magnetization as a function of the external magnetic field $H$ measured by XMCD at the Co-$L_3$ edge (778.4 eV). The measurements were conducted in a magnetic field with $\theta = 20°$.

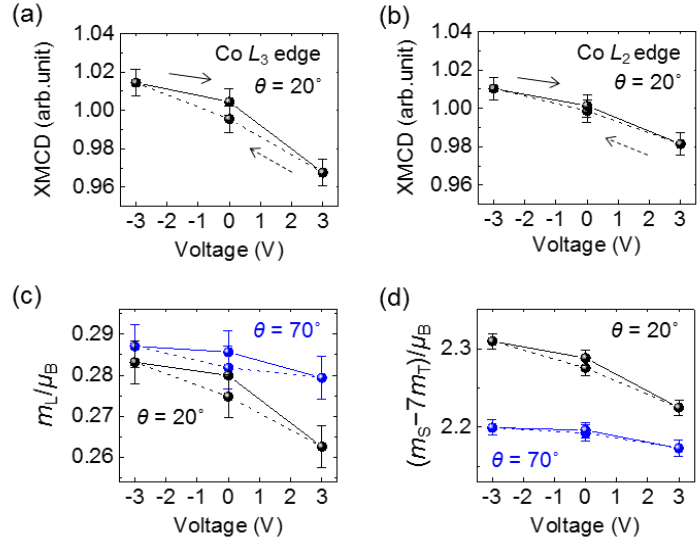

FIG. 3 (Color online) External-voltage-induced changes in XMCD at the (a) Co-$L_3$ (778.4 eV) and (b) Co-$L_2$ (793.8 eV) edges obtained using the PFY method. External-voltage-induced changes in the (c) orbital magnetic moment $m_L$ and (d) effective spin magnetic moment $m_S-7m_T$ of Co. An external voltage of ±3 V corresponds to an external electric field of ±0.2 V/nm in the 2-nm MgO dielectric. An external magnetic field of ±1.9 T was applied to saturate the magnetization of Fe/Co layer in the magnetic field direction. A negative bias voltage, where holes accumulate at the Co-MgO interface, increases both the $m_L$ and $m_S-7m_T$.

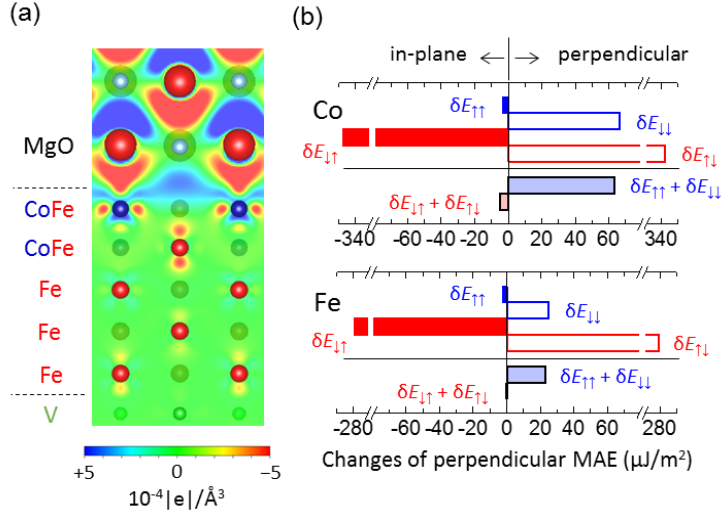

FIG. 4 (Color online) (a) Computational model with the induced change of charge density. The induced change of the charge density at an electric field of +0.2 V/nm in the MgO is subtracted from that at −0.2 V/nm. The blue and red areas represent the hole accumulation and depletion, respectively. (b) Electric-field-induced changes to the perpendicular MAE of Co and Fe atoms at the MgO interface calculated with Eq. 2. The MAE at +0.2 V/nm in the MgO is subtracted from that at −0.2 V/nm. The spin-conserved-term-induced values of the MAE change ($\delta E_{\uparrow\uparrow} + \delta E_{\downarrow\downarrow}$) in Co provide the dominant contribution to the VCMA.